\documentclass[11pt,a4paper,english]{amsart}
\usepackage{babel}
\usepackage[latin1]{inputenc}
\usepackage{amsmath}
\usepackage{amsthm}
\usepackage{amsfonts}
\usepackage{indentfirst}
\usepackage{graphicx}
\usepackage{amssymb}
\usepackage[mathscr]{eucal}
\usepackage{tikz}
\usepackage{caption}

\usepackage{enumerate}
\usepackage{amsthm}
\usepackage{amscd}
\newtheorem{teo}{Theorem}

\newtheorem{pro}[teo]{Proposition}
\newtheorem{cor}[teo]{Corollary}
\newtheorem{lem}[teo]{Lemma}

\theoremstyle{definition}
\newtheorem{rem}[teo]{Remark}
\newtheorem{de}[teo]{Definition}

\title[Steane enlargement of EAQECC]{Steane enlargement of Entanglement-Assisted Quantum Error-Correcting Codes}
\author{Carlos Galindo, Fernando Hernando and Ryutaroh Matsumoto}
\curraddr{\texttt{Carlos Galindo and Fernando Hernando:} Instituto
Universitario de Matem\'aticas y Aplicaciones de Castell\'on and
Departamento de Matem\'aticas, Universitat Jaume I, Campus de Riu
Sec. 12071 Castell\'{o} (Spain)\\
\texttt{Ryutaroh Matsumoto:} Department of Information and Communication Engineering, Tokyo Institute of Technology, Japan and Department of Mathematical Sciences, Aalborg University, 9220 Aalborg $\emptyset$, Denmark.}

\email{{\rm Galindo:} galindo@uji.es; {\rm Hernando:} carrillf@uji.es; {\rm Matsumoto:} ryutaroh@ict.e.titech.ac.jp}
\date{}

\thanks{The first two authors were partially funded by MCIN/AEI/10.13039/501100011033, by ``ERDF A way of making Europe" and by ``European Union NextGeneration EU/PRTR", grants PGC2018-096446-B-C22 and TED2021-130358B-I00, as well as by Universitat Jaume I, grant UJI-B2021-02.}


\begin{document}

\begin{abstract}
We introduce a Steane-like enlargement procedure for entanglement-assisted quantum error-correcting codes (EAQECCs) obtained by considering Euclidean inner product. We give formulae for the parameters of these enlarged codes and apply our results to explicitly compute the parameters of enlarged EAQECCs coming from some BCH codes.
\end{abstract}

\maketitle

\section{Introduction}\label{se:uno}

Quantum error-correcting codes (QECCs) are an important tool in quantum computation since they allow us to protect quantum information from quantum noise and, in particular, from decoherence. The existence of quantum processors that improve the behaviour of classical ones \cite{Aru} explains the importance of obtaining good QECCs. A $q$-ary QECC of length $n$, $q$ being a power of a prime $p$, is a subspace of the Hilbert space $\mathbb{C}^{q^n}$, where $\mathbb{C}$ denotes the complex numbers. Their parameters are written as $((n,K,d))_q$, where $K$ is the dimension and $d$ the distance. The most common QECCs are stabilizer codes \cite{Ketkar}. Many of these codes satisfy $K=q^k$, for some nonnegative integer $k$, and then their parameters are expressed as $[[n,k,d]]_q$; abusing the notation we say that $k$ is the dimension of these codes.  QECCs were  first introduced in the binary case \cite{20kkk, Gottesman, Calderbank, 7kkk, 8kkk} and later in the general case \cite{BE, AK, Ketkar, Aly, Lag2, cao-cui2, gahe, Martin-1}, where we have cited only some references of a vast  literature on the subject. QECCs in the nonbinary case are convenient for fault-tolerant quantum computation \cite{FTShor, FTKnill, FTGot, luol}.

Stabilizer quantum codes are intimately related to self-orthogonal additive codes, where duality is given by a trace-symplectic form \cite{AK, Ketkar}. As particular cases of the above construction, stabilizer quantum codes can be obtained from self-orthogonal linear codes with respect to Hermitian or Euclidean inner products. The Euclidean case corresponds to the so-called CSS construction given by Calderbank and Shor \cite{20kkk} and by Steane \cite{95kkk}. Next we recall this result \cite[Lemma 20]{Ketkar}, where $\perp_e$ stands for Euclidean dual and $\mathrm{wt}$ for Hamming weight.

\begin{teo}
\label{CSS}
Consider two $q$-ary linear codes $C_i$, $i=1,2$, with respective parameters $[n,k_i,d_i]_q$, and assume that $C_2^{\perp_e} \subseteq C_1$. Then, there exists a stabilizer quantum code with parameters $[[n, k_1+ k_2 -n, d]]_q$ with minimum distance $$d= \min \left\{ \mathrm{wt} (\boldsymbol{c}) \; | \; \boldsymbol{c} \in \left(C_1 \setminus C_2^{\perp_e}\right) \cup \left(C_2 \setminus C_1^{\perp_e}\right) \right\}.$$
\end{teo}

Quantum codes obtained from the CSS procedure enjoy several advantages. Indeed, they can be used for privacy amplification of quantum cryptography \cite{Shor-Preskill} and  for constructing asymmetric quantum codes \cite{Ioffe, Ass}, but they have mainly computational virtues, among others the smaller size of the supporting alphabet \cite{Grassl22}.


Another good property of CSS codes is that they can be improved by using the Steane enlargement procedure. This procedure was initially proposed by Steane in the binary case \cite{Steane-enl} and afterwards generalized to the general case \cite{Hamada, Chao} (see also \cite{QINP}).  Denote by $\mathbb{F}_q$ the finite field with $q$ elements, the specific result is the following one:

\begin{teo}
\label{Steane}
Let $C$ be a linear code over $\mathbb{F}_q$ of length $n$ and dimension $k$. Suppose that $C^{\perp_e} \subseteq C$ and $C$ can be enlarged to a $q$-ary linear code $C'$ of length $n$ and dimension $k' \geq k+2$. Then, there exists a stabilizer quantum code with parameters
$[[n, k+k'-n, d]]_q$, where $d \geq \min \{d_1, \lceil \frac{q+1}{q} d_2 \rceil \}$, $d_1 = \mathrm{wt} \left(C \setminus (C')^{\perp_e}\right)$ and  $d_2 = \mathrm{wt} \left(C' \setminus (C')^{\perp_e}\right)$.
\end{teo}

As mentioned, self-orthogonal codes (or codes containing their duals) have to be considered for providing stabilizer  quantum codes and this fact determines the parameters of the obtained codes. However, one can use any linear code without no condition whenever encoder and decoder share entanglement, which increases the capacity of communication \cite{Brun}. These codes are named entanglement-assisted quantum error-correcting codes (EAQECCs). Apart from the parameters length $n$, dimension $k$ and minimum distance $d$ used for  quantum codes, EAQECCs include a new one $c$, which gives the minimum number of (pairs of) maximally entangled quantum states required and then, the parameters of these codes are expressed as $[[n,k,d;c]]_q$.

A formula for computing the value $c$ of binary EAQECCs was first given for the CSS construction \cite{Hsie}. Afterwards,  always in the binary case, more general constructions where treated by Wilde and Brun \cite{Wilde}. In the general  case, formulae for obtaining the parameters of $q$-ary EAQECCs which extend those in \cite{Hsie, Wilde} can be found in \cite{QINP3}. Using them, parameters of many specific constructions of EAQECCs are recently given \cite{Qian, Sari, Chen2, Quian2, Guo, Guo2, Mesnager, GHR-RS}.

Next we recall two of the main results on EAQECCs we will use in this article. To begin with, we consider the vector space $\mathbb{F}_q^{2n}$ where $n$ is a positive integer. The {\it symplectic product} of two vectors $(\boldsymbol{x}|\boldsymbol{y})$ and $(\boldsymbol{z}|\boldsymbol{t})$ in  $\mathbb{F}_q^{2n}$ is defined as
\[
(\boldsymbol{x}|\boldsymbol{y}) \cdot_s  (\boldsymbol{z}|\boldsymbol{t}) :=   \boldsymbol{x} \cdot_e \boldsymbol{t} - \boldsymbol{z} \cdot_e \boldsymbol{y},
\]
where $\cdot_e$ means Euclidean inner product. The dual space of a vector subspace $C \subseteq \mathbb{F}_q^{2n}$ with respect to the symplectic product $\cdot_s$ is denoted by $C^{\perp_s}$.

The {\it symplectic weight} of a vector $(\boldsymbol{x}|\boldsymbol{y})$ in $\mathbb{F}_q^{2n}$ is $$\mathrm{swt} (\boldsymbol{x}|\boldsymbol{y}) := \# \{j \; | \; (x_j,y_j) \neq (0,0), 1 \leq j \leq n\},$$ $\#$ meaning cardinality and $x_j$ (respectively, $y_j$) being the $j$th coordinate of the vector $\boldsymbol{x}$ (respectively, $\boldsymbol{y}$). We define the minimum symplectic distance of a subset $S \subseteq \mathbb{F}_q^{2n}$ as
\[
d_s\left(S\right) := \min\left\{ \mathrm{swt}\left(\boldsymbol{x}|\boldsymbol{y}\right) \; \mid \;  \left(\boldsymbol{x}|\boldsymbol{y}\right) \in S \setminus \left\{ \left(\boldsymbol{0}|\boldsymbol{0}\right) \right\} \right\}.
\]

Our first result, which can be found  in \cite[Theorem 2]{QINP3}, determines the parameters of the EAQECC that one can get from a linear code $C \subseteq \mathbb{F}_q^{2n}$ over $\mathbb{F}_q$. Suppose that $C$ has dimension $n-k$. One desires to obtain a symplectic self-orthogonal $\mathbb{F}_q$-vector space $\tilde{C} \subseteq \mathbb{F}_q^{2n+2c}$, whose projection is $C$ and $c$ is the smallest number  of maximally entangled quantum states in $\mathbb{C}^q \otimes \mathbb{C}^q$. $\tilde{C}$ provides the quantum circuit which, by means of $c$ maximally entangled pairs, encodes $k+c$ logical qudits into $n$ physical qudits.

\begin{teo}\label{entang}
Let $C \subseteq \mathbb{F}_q^{2n}$ be a linear code which is generated by the rows of a matrix $(H_X|H_Z)$ of size $(n-k) \times 2n$. Then, $C$ gives rise to an EAQECC with parameters $[[n, k+c, d; c]]_q$, where
 $$2c = \mathrm{rank}\left(H_X H_Z^T - H_Z H_X^T\right)  = \dim_{\mathbb{F}_q} C - \dim_{\mathbb{F}_q} \left(C \cap C^{\perp_s} \right)$$
and $d = d_s \left(C^{\perp_s}\setminus (C\cap C^{\perp_s}) \right)$.
\end{teo}

The second mentioned result on EAQECCs \cite[Theorem 4]{QINP3} is a specialization of Theorem \ref{entang} and it shows how the CSS construction can be used for providing $q$-ary EAQECCs. To state it, consider two $\mathbb{F}_{q}$-linear codes $C_1, C_2 \subseteq \mathbb{F}_q^{n}$  of dimensions $k_1$ and $k_2$, and generator matrices $H_1$ and $H_2$, respectively. The specific result is the following one, where $d_H$ means Hamming distance, $\perp_e$ Euclidean dual and, for a matrix $M$, $M^T$ denotes its transpose.

\begin{teo}\label{entang-css}
With the above notation, the code $C_1 \times C_2 \subseteq \mathbb{F}_q^{2n}$ determines an EAQECC with parameters $[[n, n-k_1-k_2 +c, d; c]]_q$, where
\[
c = \mathrm{rank} \left(H_1H_2^T\right) =  \dim_{\mathbb{F}_{q}} C_1 - \dim_{\mathbb{F}_{q}} \left(C_1 \cap C_2^{\perp_e}\right),
\]
and
\[
d= \min \left\{ d_H\left(C_1^{\perp_e} \setminus (C_2 \cap C_1^{\perp_e})\right), d_H\left(C_2^{\perp_e} \setminus  (C_1 \cap C_2^{\perp_e})\right) \right\}.
\]
\end{teo}

In Section \ref{se:dos} of this paper we prove that, with a similar procedure to that given by Steane, one can enlarge the EAQECCs provided by Theorem \ref{entang-css} whenever the involved linear codes $C_1$ and $C_2$ coincide. Setting $C=C_1=C_2$ and expressing $C$ as a direct sum of two linear spaces $\langle B_r \rangle$ and $\langle B_t \rangle$, Theorem \ref{enlarg1} shows how to compute the parameters of the corresponding enlarged EAQECC. This enlarged code uses an invertible $t \times t$ matrix $A$, where $t$ is the dimension of the space $\langle B_t \rangle$. Under the additional hypothesis that $A$ has no eigenvalue in the supporting field and $\langle B_t \rangle \subseteq C^{\perp_e}$, Theorem \ref{case1} determines the parameters of the enlarged EAQECC. In particular, Theorem \ref{case1} proves that the enlarged code keeps the same parameter $c$ and enlarges the dimension with respect to the original EAQECC. Theorem \ref{case2}, Corollary \ref{matrix} and Remark \ref{La13} study and show the advantages produced by the Steane enlargement of EAQECCs obtained when the spaces $\langle B_r \rangle$ and $\langle B_t \rangle$ are Euclidean orthogonal.

Our procedure can be carried out for any decomposition of a linear code as mentioned. As a specific case, Section \ref{se:tres} explains how to put our results into practice through certain BCH codes. Regarding BCH codes as subfield-subcodes and applying Theorem \ref{case1} and  previous results, Theorems \ref{BCHcase1} and \ref{Reciprocal}  state explicitly those parameters of the Steane enlargement of the EAQECCs given by suitable BCH codes. Finally, in Theorems \ref{BCHcase2} and \ref{Reciprocal-2} and Remark \ref{La22}, by considering another families of BCH codes, we determine the corresponding parameters of EAQECCs deduced from Theorem \ref{case2}, Corollary \ref{matrix} and Remark \ref{La13}.

In this section we have given a brief introduction to QECCs and EAQECCs, recalling some of the main known results which will be used later. The goal of this article is to explain how a Steane enlargement of EAQECCs can be achieved. We also specialize this construction to certain BCH codes and, as a consequence, we obtain new EAQECCs which enjoy interesting computational advantages (see the last part of the paragraph after Theorem \ref{CSS}). Section \ref{se:dos} introduces the Steane enlargement of EAQECCs and proves several results about their parameters, while Section \ref{se:tres} applies the results in Section \ref{se:dos} to compute the parameters of enlarged codes of EAQECCs associated to BCH codes. Some good codes deduced from our procedure are also presented at the end of this last section.


\section{Steane enlargement of EAQECCs} \label{se:dos}
The goal of this section is to prove that a procedure like the Steane enlargement  showed in Theorem \ref{Steane} can be used in the framework of EAQECCs. Firstly we will see how it applies to codes as described in Theorem \ref{entang-css} and then we will show that, under certain conditions, we are able of determine the parameters of the enlarged EAQECCs and how they improve the original ones.

Let $C \subseteq  \mathbb{F}_q^{n}$  be an $\mathbb{F}_q$-linear code with parameters $[n,k,\delta]_q$. Let $B$ be a matrix with entries in  $\mathbb{F}_q$ whose rows are linearly independent vectors of $\mathbb{F}_q^n$. Along this paper, by convenience, we denote by $\langle B \rangle$ the vector subspace of $\mathbb{F}_q^{n}$ generated by the rows of $B$. Assume that $C= \langle B_r \rangle \oplus \langle B_t \rangle$, where $B_r$ (respectively, $B_t$) are $r \times n$ (respectively, $t \times n$) generator matrices of the $\mathbb{F}_q$-linear subcodes $\langle B_r \rangle$ (respectively, $\langle B_t \rangle$) of $C$, and where $r = \dim_{\mathbb{F}_q} \langle B_r \rangle $ and $t = \dim_{\mathbb{F}_q} \langle  B_t \rangle $. Applying Theorem \ref{entang-css} for $C_1= C_2 =C$, one obtains an EAQECC, $\tilde{C}$, with parameters
\begin{equation}
\label{inicial}
[[n, n- 2k +c, d;c]]_q,
\end{equation}
where $k=r+t$, $d=d_H (C^{\perp_e} \setminus C)$ and
\begin{eqnarray}
 c = \mathrm{rank}\left[
    \left(
    \begin{array}{c} B_r \\ B_t \end{array} \right)
    ( B_r^T B_t^T)
    \right]
  &=&
  \mathrm{rank}\left[
    \left(
    \begin{array}{cc}
      B_r B_r^T & B_r B_t^T\\
      B_t B_r^T & B_t B_t^T
    \end{array} \right)
    \right]. \label{thec}
\end{eqnarray}

Next, we introduce the linear code $D_A$ we desire to use in order to provide the Steane enlargement of $\tilde{C}$.

\begin{de}
\label{D}
 Given a code $C= \langle B_r \rangle \oplus \langle B_t \rangle $ as above and such that $\dim_{\mathbb{F}_q} \langle B_t \rangle \geq 2$, we define the code $D_A$ as the $\mathbb{F}_q$-linear code $D_A \subseteq \mathbb{F}_q^{2n}$ whose generator matrix is
\begin{equation}
  \left(
  \begin{array}{cc}
    B_t & A B_t\\
    Br & 0 \\
   0 & B_r
  \end{array}\right), \label{eq1}
\end{equation}
where $A$ is a $t \times t$ invertible matrix over $\mathbb{F}_q$.
\end{de}

Our general result on Steane enlargement is the following one:

\begin{teo}
\label{enlarg1}
Let $C= \langle B_r \rangle \oplus \langle B_t \rangle \subseteq  \mathbb{F}_q^{n}$ be an $\mathbb{F}_q$-linear code such that $\dim_{\mathbb{F}_q} \langle B_t \rangle \geq 2$. Assume that $A$ is a $t \times t$ invertible matrix over $\mathbb{F}_q$ which has no eigenvalue in $\mathbb{F}_q$. Then $D_A$ gives rise to an EAQECC, $\tilde{D}_A$, which is a Steane enlargement of the EAQECC, $\tilde{C}$, with parameters
$$
[[n, n-2r-t +c', d', c']]_q,
$$
where
$d' \geq \min \left\{ \delta_1, \left\lceil \left( 1+ \frac{1}{q}\right)  \delta_2 \right\rceil  \right\}$, $\delta_1 = d_H (C^{\perp_e})$ and  $\delta_2 = d_H \left(\langle B_r \rangle^{\perp_e}\right)$, and
\begin{equation}
    c' = \frac{1}{2} \mathrm{rank} \left(
  \begin{array}{ccc}
    B_t B_t^T A^T - A B_t B_t^T  & - A B_t B_r^T &  B_t B_r^T \\
    B_r B_t^T A^T & 0 & B_r B_r^T \\
    -B_r B_t ^T  & -B_r B_r^T & 0
  \end{array}
  \right).\label{eq5}
  \end{equation}
\end{teo}

\begin{proof}
The proof follows by applying Theorem \ref{entang} to the linear code $D_A \subseteq \mathbb{F}_q^{2n}$. The size of its generator matrix  is $(n-(n-2r-t)) \times 2n$. By Theorem \ref{entang}, the dimension of the obtained EAQECC is $n - 2r - t +c'$ and the number of maximally entangled pairs $c'$ is given by the formula
\begin{eqnarray*}
  2c' \; = \; \mathrm{rank} \left[
    \left(
    \begin{array}{c} B_t \\ B_r \\ 0 \end{array} \right)
    \left( B_t^T A^T \;\; 0  \;\; B_r^T \right) \right.
    & - &
   \left. \left(
    \begin{array}{c} A B_t \\ 0 \\ B_r \end{array} \right)
    \left( B_t^T \;\;  B_r^T \; \; 0 \right)
    \right].
    \end{eqnarray*}

Then, we have proved that, with the exception of the minimum distance, the parameters of the Steane enlargement of the EAQECC, $\tilde{C}$, are as in the statement.

With respect to the minimum distance $d'$, we are going to prove that $$d_s \left(D_A^{\perp_s} \setminus D_A \cap D_A^{\perp_s}\right) \geq \min \left\{ \delta_1, \left\lceil \left( 1+ \frac{1}{q}\right)  \delta_2 \right\rceil  \right\},$$
which, again by Theorem \ref{entang}, finishes our proof.

Denote by $G$ a generator matrix of the Euclidean dual $C^{\perp_e}$ with size $(n-(r+t)) \times n$ and set $\left( G' \;|\; G \right)^T$ a generator matrix of the dual vector space $\langle  B_r \rangle^{\perp_e}$. Then, by the proof of Theorem 2.6 of \cite{Chao} (see also \cite{Hamada}),  the symplectic dual $D_A^{\perp_s}$ of $D_A$ has
\begin{equation*}
  \left(
  \begin{array}{cc}
   \bar{A} G' & G'\\
   G & 0 \\
   0 & G
  \end{array}\right),
\end{equation*}
as a generator matrix, where $\bar{A} := G' B_t^T \left( A^T \right)^{-1} \left(G' B_t^T \right)^{-1}$.

Now, $\bar{A}$ has no eigenvalue in $\mathbb{F}_q$ because $A$ has no eigenvalue in $\mathbb{F}_q$ and if an invertible matrix $M$ with entries in $\mathbb{F}_q$ has an eigenvalue $\lambda \in \mathbb{F}_q$, then  $M^T$, $M^{-1}$ and $PMQ$, where $P$ and $Q$ are invertible matrices with entries in $\mathbb{F}_q$, also have an eigenvalue in $\mathbb{F}_q$. To conclude, again by  the proof of Theorem 2.6 of \cite{Chao},
$$d_s \left( D_A^{\perp_s} \setminus D_A \cap D_A^{\perp_s} \right) \geq \min \left\{ d_H \left( \langle G \rangle\right),  d_H\left( \langle B_r \rangle^{\perp_e} \right) \right\},
$$
proving the lower bound for $d'$ in the statement.
\end{proof}

The following subsections study specific cases where one can give more information about the advantages of using Steane enlargement of EAQECCs. An application of these results by considering BCH codes will be given in Section \ref{se:tres}.

\subsection{The case when $\langle B_t \rangle$ and $C$ are Euclidean orthogonal}
\label{sub:one}
In this subsection we consider the Steane enlargement $\tilde{D}_A$ of the EAQECC, $\tilde{C}$, defined by an $\mathbb{F}_q$-linear code $C= \langle B_r \rangle \oplus \langle B_t \rangle$ such that $\langle B_t \rangle \subseteq C^{\perp_e}$. Keeping the above notation, our result is the following one.

\begin{teo}
\label{case1}
Let $C= \langle B_r \rangle \oplus \langle B_t \rangle \subseteq \mathbb{F}_q^n$ be an $\mathbb{F}_q$-linear code such that $\langle B_t \rangle \subseteq C^{\perp_e}$ and $\dim_{\mathbb{F}_q} \langle B_t \rangle \geq 2$. Set $t = \dim_{\mathbb{F}_q} \langle B_t \rangle$ and assume that $A$ is a $t \times t$ invertible matrix over $\mathbb{F}_q$ which has no eigenvalue in $\mathbb{F}_q$. Denote by $c$ the minimum required number of maximally entangled states in the EAQECC, $\tilde{C}$, determined by $C$. Then, the linear code $D_A$, introduced in Definition \ref{D}, gives rise to an  EAQECC, $\tilde{D}_A$, which is a Steane enlargement of  $\tilde{C}$, with parameters
$$
[[n, n-2r-t +c, d', c]]_q,
$$
where
$d' \geq \min \left\{ \delta_1, \left\lceil \left( 1+ \frac{1}{q}\right)  \delta_2 \right\rceil  \right\}$, $\delta_1 = d_H \left(C^{\perp_e}\right)$ and  $\delta_2 = d_H \left(\langle B_r \rangle^{\perp_e}\right)$.
\end{teo}

\begin{proof}
 Theorem \ref{enlarg1} determines the parameters  given in the statement of the EAQECC $\tilde{D}_A$ with the exception of the fact that the minimum required number of maximally entangled states $c'$ in $\tilde{D}_A$ equals $c$. Now, $\langle B_t \rangle \subseteq C^{\perp_e}$ and then $B_t B_t^T =0$ and $B_r B_t^T =0$. Then, by (\ref{thec}),
 \begin{equation*}
    c =  \mathrm{rank} \left(
  \begin{array}{cc}
      B_r B_r^T & 0\\
      0 & 0
    \end{array} \right)
    =  \mathrm{rank} \left( B_r B_r^T \right).
  \end{equation*}

Finally, by (\ref{eq5}) and the above equalities, it holds that
\begin{equation*}
    c' = \frac{1}{2} \mathrm{rank} \left(
  \begin{array}{ccc}
    0  & 0 &  0 \\
    0 & 0 & B_r B_r^T \\
   0  & -B_r B_r^T & 0
  \end{array}
  \right) = \mathrm{rank} \left( B_r B_r^T \right) = c,
  \end{equation*}
 which concludes the proof.
\end{proof}

The above result shows that suitable choices of linear codes $C$ allow us to get Steane enlargements of the EAQECC given by $C$ that enlarge its dimension (by $t$) and keep the number $c$ of required maximally entangled states.

To finish this subsection we introduce a class of matrices which allows us to get matrices $A$ as required in Theorems \ref{enlarg1} and \ref{case1}. This class of matrices will be also used in the forthcoming Subsection \ref{sub:two}. Let $j \geq 2$ be an integer, consider the monic polynomial
\begin{equation}
\label{polin}
h_{\boldsymbol{a}} (X) := X^j + a_{j-1} X^{j-1} + \cdots + a_1 X + a_0 \in \mathbb{F}_q [X]
\end{equation}
and its corresponding  $j \times j$ (companion) matrix with entries in $\mathbb{F}_q$:
\[
L_j (h_{\boldsymbol{a}})= \left(
\begin{array}{cccccc}
  0 & 1 & 0 & \cdots & 0 & 0\\
  0 & 0 & 1 & \cdots & 0 & 0\\
  \vdots & \vdots & \ddots & \ddots & \cdots & \vdots \\
  \vdots & \vdots & \vdots & \ddots & \ddots & \vdots \\
  0 & 0 & 0 & \cdots & 0 & 1\\
  -a_0 & -a_1 & -a_2 & \cdots & -a_{j-2} & -a_{j-1}
\end{array}
\right).
\]
Then the following proposition holds.
\begin{pro}
$h_{\boldsymbol{a}} (X)$ is the characteristic polynomial of the matrix $L_j (h_{\boldsymbol{a}})$.
\end{pro}
\begin{proof}
A proof can be found in \cite[Lemma 7]{Hamada} or in \cite[Lemma 2.5]{Chao}.
\end{proof}

\begin{rem}
\label{rem}
{\rm
Let $t \geq 2$ be a positive integer. Consider the polynomial $\bar{h}_t:=X^t + X^{t-1} \in \mathbb{F}_q [X]$ which gives rise to the map $\varphi_{\bar{h}_t}: \mathbb{F}_q \rightarrow \mathbb{F}_q $, $\varphi_{\bar{h}_t}(x) = \bar{h}_t (x)$. Clearly $\varphi_{\bar{h}_t}$ is not one-to-one and there exists $0 \neq \xi \in \mathbb{F}_q$ which is not in the image of $\varphi_{\bar{h}_t}$. Then the polynomial $h_t := \bar{h}_t -\xi \in \mathbb{F}_q [X]$ has no roots in $\mathbb{F}_q$. As a consequence $L_t(h_t)$ is a suitable choice of a matrix $A$ for Theorem \ref{case1}.}
\end{rem}

\subsection{The case when $\langle B_r \rangle$ and $\langle B_t \rangle$ are Euclidean orthogonal}
\label{sub:two}
This subsection studies the Steane enlargement of the EAQECC $\tilde{C}$ given by a code $C= \langle B_r \rangle \oplus \langle B_t \rangle \subseteq \mathbb{F}_q^n$ satisfying that the codes $\langle B_r \rangle$ and $\langle B_t \rangle$ are Euclidean orthogonal. Note that Subsection \ref{sub:one} studies a special situation of the present subsection, where we are going to give the parameters of the Steane enlargement  $\tilde{D}_A$. In this case, additional conditions will be required to the matrix $A$.

Set $ \langle  B_t \rangle = \langle B_{t_\ell} \rangle  \oplus \langle B_{t_Q} \rangle$, where $\langle B_{t_Q} \rangle  \subseteq \langle B_{t} \rangle^{\perp_e} $. We denote by $t_\ell$ the dimension of the linear code $\langle B_{t_\ell} \rangle$. Without loss of generality, assume that the rows of $B_{t_\ell}$ are compatible with a geometric decomposition of $\mathbb{F}_q^n$ (see \cite{Diego}) and by \cite[Section 2.4]{QINP3}, $Z:=B_{t_\ell} B_{t_\ell}^T$ is a matrix such that all its elements are zero but diagonal boxes of the form
\[
  \left(
  \begin{array}{cc}
    0 & 1\\
  1 & 0
  \end{array}\right),
\]
or $(z_{i})$, $z_{i} \neq 0$, which in characteristic $2$ may include a box as follows:
\[
  \left(
  \begin{array}{cc}
    0 & 1\\
  1 & 1
  \end{array}\right).
\]
That is,
\[
Z = \left(
\begin{array}{l@{\hspace{0.4cm}}l@{\hspace{0.4cm}}l@{\hspace{0.4cm}}l@{\hspace{0.4cm}}l
@{\hspace{0.4cm}}l@{\hspace{0.4cm}}l@{\hspace{0.4cm}}l@{\hspace{0.4cm}}l @{\hspace{0.4cm}}
l@{\hspace{0.4cm}}}

0 & 1 & \nonumber  & \nonumber & \nonumber  & \nonumber & \nonumber & \nonumber & \nonumber & \nonumber  \\

1 & 0 & \nonumber  & \nonumber & \nonumber  & \nonumber & \nonumber & \nonumber & \nonumber & \nonumber \\

\nonumber  & \nonumber & \ddots  & \nonumber & \nonumber  & \nonumber & \nonumber   & \nonumber & \nonumber  & \nonumber  \\

\nonumber  & \nonumber & \nonumber  & 0  & 1 &  \nonumber & \nonumber & \nonumber  & \nonumber  & \nonumber \\

\nonumber  & \nonumber &  \nonumber & 1  & 0 & \nonumber  & \nonumber & \nonumber & \nonumber & \nonumber \\

\nonumber  & \nonumber &  \nonumber & \nonumber  & \nonumber & z_1  & \nonumber  &  \nonumber & \nonumber  & \nonumber   \\

\nonumber  & \nonumber &  \nonumber & \nonumber  & \nonumber & \nonumber  & \ddots &  \nonumber & \nonumber  & \nonumber   \\

\nonumber &  \nonumber & \nonumber  & \nonumber & \nonumber  & \nonumber & \nonumber  & z_s & \nonumber & \nonumber\\

 \nonumber &  \nonumber & \nonumber  & \nonumber & \nonumber  & \nonumber & \nonumber & \nonumber & 0 & 1  \\

 \nonumber &  \nonumber & \nonumber  & \nonumber & \nonumber  & \nonumber & \nonumber & \nonumber & 1 & 1
\end{array}
\right),
\]
where the last box may only appear in characteristic $2$.

Now consider an invertible matrix
\[
A = \left(
\begin{array}{cc}
  A_0 Z^{-1} & 0 \\
  0 & A_1
\end{array}
\right),
\]
where $A_0$ (respectively, $A_1$) are  $t_\ell  \times t_\ell$ (respectively, $(t- t_\ell)  \times (t- t_\ell))$ matrices with entries in  $\mathbb{F}_q$. We also assume that $A$ has no eigenvalue in the finite field  $\mathbb{F}_q$. Then, we are ready to state our main result in this subsection.

\begin{teo}
\label{case2}
Let $C= \langle B_r \rangle \oplus \langle B_t \rangle \subseteq \mathbb{F}_q^n$ be an $\mathbb{F}_q$-linear code as before. Denote by $c$ the minimum required number of maximally entangled states in the EAQECC $\tilde{C}$ determined by $C$. Then, the linear code $D_A$ gives rise to an EAQECC $\tilde{D}_A$, which is a Steane enlargement of $\tilde{C}$, with parameters
$$
[[n, n-2r-t +c', d', c']]_q,
$$
where
\[
c' = (c-t_\ell) + \frac{1}{2} \mathrm{rank} \left(A_0 - A_0^T\right) = c- \mathrm{rank}\left(B_t B_t^T\right) + \frac{1}{2} \mathrm{rank} \left(A_0 - A_0^T\right)
\]
and
$d' \geq \min \left\{ \delta_1, \left\lceil \left( 1+ \frac{1}{q}\right)  \delta_2 \right\rceil  \right\}$, $\delta_1 = d_H \left(C^{\perp_e}\right)$ and  $\delta_2 = d_H \left(\langle B_r \rangle^{\perp_e}\right)$.
\end{teo}

\begin{proof}
Theorem \ref{enlarg1} shows the parameters in the statement with the exception of the formula for $c'$. The number of maximally entangled states $c'$ depends on the rank of the matrix in (\ref{eq5}). We have assumed that the codes $\langle B_r \rangle$ and $\langle B_t \rangle$ are Euclidean orthogonal, therefore $B_rB_t^T=0$ and the boxes in positions $(1,2)$, $(1,3)$, $(2,1)$ and $(3,1)$ of the matrix in (\ref{eq5}) vanish. In addition, recalling that $Z= B_{t_\ell} B_{t_\ell}^T$ is an invertible square matrix of size $t_\ell$, it holds that
 \begin{equation*}
    B_t B_t^T =   \left(
  \begin{array}{cc}
      Z & 0\\
      0 & 0
    \end{array} \right)
  \end{equation*}
and then, the box in position $(1,1)$ of the matrix in  (\ref{eq5}), which is $B_t B_t^T A^T - A B_t B_t^T$, is equal to
\[
 \left( \begin{array}{cc}
      Z & 0\\
      0 & 0
    \end{array} \right) \left( \begin{array}{cc}
      (Z^{-1})^T A_0^T & 0\\
      0 & A_1^T
    \end{array} \right) - \left( \begin{array}{cc}
     A_0 Z^{-1} & 0\\
      0 & A_1
    \end{array} \right) \left( \begin{array}{cc}
      Z & 0\\
      0 & 0
    \end{array} \right) = \left( \begin{array}{cc}
     A_0^T -A_0  & 0\\
      0 & 0
    \end{array} \right).
\]

Therefore, it holds that
\begin{equation}
\label{AA}
c' =  \frac{1}{2} \left( \mathrm{rank} \left(A_0 -A_0^T\right) \right) + \mathrm{rank} \left( B_r B_r^T\right).
\end{equation}
Now, taking into account that $\langle B_r \rangle$ and $\langle B_t \rangle$ are Euclidean orthogonal, by Equality (\ref{thec}), we get
\begin{equation}
\label{BB}
c= \mathrm{rank} \left( B_r B_r^T \right) + \mathrm{rank} \left( B_t B_t^T \right) = \mathrm{rank} \left( B_r B_r^T \right) +t_\ell.
\end{equation}
Combining equalities (\ref{AA}) and (\ref{BB}), the equalities for $c'$ in the statement are proved.
\end{proof}

We finish this subsection by showing that a matrix $A$ given in terms of the above matrices $L_j$ is suitable for our purposes.

\begin{lem}
\label{matL}
Let $h_{\boldsymbol{a}} (X)$ be a polynomial as in (\ref{polin}) and consider the companion matrix $L_j := L_j (h_{\boldsymbol{a}})$.  Then $$\mathrm{rank} (L_j -L_j^T) \geq j- 2.$$ Moreover, $\mathrm{rank} (L_j -L_j^T)=j-1$ if $j$ is odd; otherwise, $\mathrm{rank} (L_j -L_j^T)=j$ if and only if
\[
1 + a_0 + a_2 + a_4 + \cdots +a_{j-2} \neq 0.
\]
\end{lem}
\begin{proof}
For a start, it holds that
\[
L_j -L_j^T
\]
\[ = \left(
\begin{array}{cccccccccc}
  0 & 1 & 0 & 0& 0& \cdots & 0 & 0 &0 & a_0\\
  -1 & 0 & 1 & 0& 0& \cdots & 0 & 0 & 0 & a_1\\
  0 & -1 & 0 & 1& 0& \cdots & 0 & 0 & 0 & a_2\\
  \vdots & \vdots & \vdots & \vdots & \vdots &  \cdots & \vdots & \vdots &  \vdots  & \vdots \\
  0 & 0 & 0 & 0& 0& \cdots & -1 & 0 & 1 & a_{j-3}\\
  0 & 0 & 0 & 0& 0& \cdots & 0 & -1 & 0 & 1 + a_{j-2}\\
  -a_0 &   -a_1 &   -a_2 &   -a_3 &   -a_4 & \cdots & -a_{j-4} & -a_{j-3} & -1 - a_{j-2} & 0
\end{array}
\right).
\]

The matrix $L_j -L_j^T$ is skew-symmetric. This matrix is also alternate even in characteristic $2$ because the elements of its diagonal vanish. Then its rank is even \cite{Delgo}. In addition, if one deletes the last two rows and the first and the last column, one gets a triangular matrix with ones in the diagonal. Then $\mathrm{rank} (L_j -L_j^T) \geq j- 2$ and we have proved our first statement. Since the rank is even, the statement that $\mathrm{rank} (L_j -L_j^T) = j-1$ when $j$ is odd is clear.

Finally assume that $j$ is even. Then,  $\mathrm{rank} (L_j -L_j^T)$ equals either $j$ or $j-2$. We are going to prove that this rank is $j$ if and only if $\det (L_j -L_j^T) \neq 0$ if and only if $(1 + a_0 + a_2 + a_4 + \cdots +a_{j-2}) ^2 \neq 0$ if and only if $(1 + a_0 + a_2 + a_4 + \cdots +a_{j-2})  \neq 0$, which concludes the proof.
Thus, the fact to prove is that, for $j$ even,
\[
\det (L_j -L_j^T) = (1 + a_0 + a_2 + a_4 + \cdots +a_{j-2}) ^2.
\]
Indeed, adding the odd rows to the  $(j-1)$th row, we get a new matrix with the same determinant and the same rows with the exception of the $(j-1)$th one, all the entries of this $(j-1)$th row are zeros with the exception of the last one, which is
\[1+a_0+a_2+a_4+ \cdots + a_{j-2}.\]
Now we use the Laplace expansion along the $(j-1)$th row, getting a unique non-vanishing summand. Next we again perform  Laplace expansions using, successively, those rows having a unique one as entry (these ones are in the odd files of the initial matrix). Then, it remains to compute the minor given by the determinant of the matrix
\[
\left(
\begin{array}{cccccccc}
  -1 & 1 & 0 &  0& \cdots & 0 & 0 &0 \\
  0 & -1 & 1 & 0&  \cdots & 0 & 0 & 0 \\
  0 & 0 & -1 & 1& \cdots & 0 & 0 & 0 \\
  \vdots & \vdots & \vdots & \vdots &   \cdots  & \vdots & \vdots  & \vdots \\
  0 & 0 & 0 & 0&  \cdots & 0 & -1 & 1\\
  -a_0 &   -a_2 &   -a_4 &   -a_6 &  \cdots & -a_{j-6} & -a_{j-4} & -1 - a_{j-2}
\end{array}
\right).
\]
Finally, adding all the columns in the above matrix to the last column, we get the following matrix (with the same determinant):
\[
\left(
\begin{array}{cccccccc}
  -1 & 1 & 0 &  0& \cdots & 0 & 0 &0 \\
  0 & -1 & 1 & 0&  \cdots & 0 & 0 & 0 \\
   0 & 0 & -1 & 1& \cdots & 0 & 0 & 0 \\
\vdots & \vdots & \vdots & \vdots &   \cdots  & \vdots & \vdots  & \vdots \\
  0 & 0 & 0 & 0&  \cdots & 0 & -1 & 0\\
  -a_0 &   -a_2 &   -a_4 &   -a_6 &  \cdots & -a_{j-6} & -a_{j-4} & -(1+a_0+a_2+ \cdots + a_{j-2})
\end{array}
\right).
\]
This proves that $\det (L_j -L_j^T) = \left(1+a_0+a_2+a_4+ \cdots + a_{j-2}\right)^2$, and it finishes the proof.
\end{proof}

Keep the notation as in Remark \ref{rem}. One can use the matrix $ L_{t-t_\ell} \left( h_{t-t_\ell} \right)$ as the box $A_1$ in the matrix $A$ considered in Theorem \ref{case2}. 

We look for matrices $A_0$ of the type $L_j$ to guarantee that $A$ has no eigenvalue in $\mathbb{F}_q$. Our next proposal allows us to get matrices of the mentioned type such that $\mathrm{rank}(A_0-A_0^T)$ is maximum. Note that, by Theorem \ref{case2}, this fact enlarges the entanglement but also the dimension of the obtained codes.

To begin with, suppose that either the characteristic of the field $\mathbb{F}_q$ is odd or it is even and $t_\ell >2$. Then, set $\tilde{\phi}_{t_\ell}:=X^{t_\ell} - X^{t_\ell -2} \in \mathbb{F}_q [X]$. The attached map $$\varphi_{\tilde{\phi}_{t_\ell}}: \mathbb{F}_q  \rightarrow \mathbb{F}_q$$ is $\varphi_{\tilde{\phi}_{t_\ell}}(x) = \tilde{\phi}_{t_\ell} (x)$.
As in Remark \ref{rem}, $\varphi_{\tilde{\phi}_{t_\ell}}$ is not bijective and one can find a nonzero element $\xi_1$ in $\mathbb{F}_q $ such that the polynomial $\phi_{t_\ell} := \tilde{\phi}_{t_\ell} - \xi_1 \in \mathbb{F}_q [X]$
has no roots in $\mathbb{F}_q$. In this case,  $L_{t_\ell} (\phi_{t_\ell})$ is a suitable matrix for being the matrix $A_0$ involved in the box $(1,1)$ of $A$.

When $q \neq 2$, the characteristic of the field $\mathbb{F}_q$ is two and $t_\ell =2$, it suffices to consider the matrix $L_{t_2} (\phi_{2})$ defined by $\tilde{\phi}_{2} = \bar{h}_2$, where $\bar{h}_2$ is the polynomial given in Remark \ref{rem}, and consider $\phi_{2} = \tilde{\phi}_{2} - \xi$, where $\xi \neq 0, 1$ is a suitable value in $\mathbb{F}_q$. Notice that, in this case, a polynomial $X^2+X-\varsigma$, $\varsigma \in \mathbb{F}_q$,  either has two different roots or it is irreducible.

\begin{cor}
\label{matrix}
Let $A_0$ and $A_1$ be matrices as described in the above paragraphs. Then, the matrix
\[
A = \left(
\begin{array}{cc}
  A_0 Z^{-1} & 0 \\
  0 & A_1
\end{array}
\right)
\]
can be used to provide a Steane enlargement $\tilde{D}_A$ of  the EAQECC  $\tilde{C}$  given in Theorem \ref{case2}. Furthermore, the rank of the matrix $A_0 - A_0^T$ is as large as  possible. That is to say, it is $t_\ell$ (respectively, $t_\ell -1$) whenever $t_\ell$ is even (respectively, odd).
\end{cor}
\begin{proof}
The proof follows from the fact that our choice of $A_0$ and $A_1$ implies that they have no eigenvalue in $\mathbb{F}_q$ and, then, the same happens to $A$. Finally by Lemma \ref{matL},  if $t_\ell$ is even, the value $1+a_0+a_2+a_4+ \cdots + a_{j-2}$ corresponding to $\phi_{t_\ell}$ is $-\xi_1 \neq 0$ and its rank is maximum.
\end{proof}

\begin{rem}
\label{La13}
{\rm When $t_\ell$ is odd, $\mathrm{rank}(A_0-A_0^T)=t_\ell -1$ is the only possibility if one considers matrices $A_0$ of the type $L_j$.

Assume now that $t_\ell$ is even. If $t_\ell \geq q$ and the characteristic of $\mathbb{F}_q$ is two, we consider the polynomial $\eta_{t_\ell} := X^{t_\ell} - X^{t_\ell- q +1}-1$.  Assume that $q$ is odd. Soon we will show that there exists $0 \neq a \in \mathbb{F}_q$ such that the polynomial $X^2+aX-1 \in \mathbb{F}_q[X]$  has no roots in $\mathbb{F}_q$. As a consequence, the polynomial $\eta_{2w} = X^{2w}+aX^{w}-1$, $w$ being an odd positive integer, has no roots in $\mathbb{F}_q$. Therefore, if $t_\ell \geq q$ when the characteristic is two and, otherwise, when  $t_\ell = 2w$, the polynomials $\eta_{t_\ell}$ provide matrices $A_0 := L_{t_\ell} (\eta_{t_\ell})$ which are suitable for our purposes and satisfy  $\mathrm{rank}(A_0-A_0^T)=t_\ell -2$. For the remaining cases, we have no generic candidate for $A_0$ such that $\mathrm{rank}(A_0-A_0^T)=t_\ell -2$, however for specific cases and moderate values of $q$ and $t_\ell$, it is not hard to find suitable polynomials $\eta_{t_\ell}$ and attached matrices $A_0$ with the above mentioned rank. Note also that the propagation rules stated in \cite{LuoEz} does not work here since our codes do not come from the Hermitian construction.

It remains to prove that $X^2+aX-1$ is irreducible for some $a \neq 0$. It holds if and only if there exists  $0 \neq a \in \mathbb{F}_q$  such that $a^2 + 4 \neq b^2$ for all $b \in \mathbb{F}_q$. Consider the attached equation (1): $x^2 + 4 = y^2$, $x, y \in \mathbb{F}_q$ and new variables $x_1 = (x+y)/2$ and $y_1 = (y-x)/2$ which satisfy $x=x_1 - y_1$ and $y=x_1 + y_1$. It follows that $(x,y)$ is a solution of (1) if and only if $(x_1,y_1)$ is a solution of the equation (2): $x_1 y_1 =1$. Thus the map $\zeta: \mathbb{F}_q^* \rightarrow \mathbb{F}_q$ defined $\zeta(x_1) =x_1 - x_1^{-1} = x_1^{-1} (x_1^2 -1)$ takes the solutions of (2) into the solutions of (1) and $x_1 =1$ and $x_1 = -1$ give solutions of (1) with $x=0$. This proves that there is some value $a \neq 0$ in $\mathbb{F}_q$ satisfying that $a^2 + 4 = y^2$ has no solution in $\mathbb{F}_q$, which concludes the proof.
}
\end{rem}

\section{Steane enlargement of EAQECCs given by BCH codes}
\label{se:tres}

In this last section we study the Steane enlargement of EAQECCs given by some BCH codes. We divide it in two subsections. The first one shows our results while the second one gives a few examples.

\subsection{Results}
\label{results}

BCH codes are cyclic codes but we prefer to regard them as subfield-subcodes of certain evaluation codes \cite{Bier, Cas}. In fact our BCH codes are $J$-affine variety codes in one variable with $J=\{1\}$ as introduced in \cite{QINP2}, and we will use some results from this source.

Set $q=p^m$, $m \geq 1$ and $n= p^m -1$. We consider the evaluation map
$$
\mathrm{ev}: \frac{\mathbb{F}_q [X]}{\langle X^n - 1\rangle} \rightarrow \mathbb{F}_{q}^{n},
$$
given by $\mathrm{ev}(h) = (h(R_1), \ldots, h(R_n))$, where $\{R_i\}_{i=1}^n$ is the set of $n$th roots of unity in the finite field $\mathbb{F}_q$. Now define $H:=\{0, 1, \ldots, n-1\}$ and, for sets $\emptyset \neq \Delta \subseteq H$, denote by $C_\Delta$ the linear code over $\mathbb{F}_q$ generated by $\{\mathrm{ev}(X^i) : i \in \Delta\}$. Consider also a positive integer $s$ such that $s \neq m$ divides $m$ and then, BCH codes over $\mathbb{F}_{p^s}$ are subfield-subcodes of codes $C_\Delta$, that is codes the form $\mathbb{F}_{p^s}^n \cap C_\Delta$. Given $a \in H$, a minimal cyclotomic coset  (with respect to $n$ and $s$) is a set of elements  $\mathcal{I}_a :=\{a p^{\ell s} : \ell \geq 0\}$, where the products are carried out modulo $n$ (i.e., both $a$ and the elements in $\mathcal{I}_a$ are representatives in $H$ of classes in the congruence ring $\mathbb{Z}_n$). Denote by $i_a$ the cardinality of $\mathcal{I}_a$ and set $\mathcal{I}_a^R = \mathcal{I}_{n-a}$, which we name the reciprocal coset of $\mathcal{I}_a$. Moreover, $\mathcal{I}_a$ is named symmetric when $\mathcal{I}_a =\mathcal{I}_a^R$ and, otherwise, it is called to be asymmetric.

Within each minimal cyclotomic coset, we take its minimal element for the natural ordering and denote by $\mathcal{A}= \{a_0=0 < a_1 < \cdots < a_z\}$ the set of these minimal representatives. Then $\left\{ \mathcal{I}_{a_\nu} \right\}_{\nu=0}^z$ is the set of minimal cyclotomic cosets (with respect to $n$ and $s$). We will use the following result which can be found in \cite[propositions 1 and 2]{GHR-RS}.
\begin{pro}
\label{grado2}
Keep the above notation.
\begin{enumerate}
\item Assume that $\Delta=\cup_{\nu=\ell_1}^{\ell_2} \mathcal{I}_{a_\nu}$, where $\ell_1 <\ell_2$ are in $\{0, 1, \ldots,  z\}$. Then, the dimension of the subfield-subcode  $\mathbb{F}_{p^s}^n \cap C_\Delta$ is equal to $\sum_{\nu =\ell_1}^{\ell_2} i_{a_\nu}$. The same result is true when $\Delta$ is union of nonconsecutive minimal cyclotomic cosets.
\item Assume now that $\ell_1=0$ and denote by $\delta$ the minimum distance of the Euclidean dual code of $\mathbb{F}_{p^s}^n \cap C_\Delta$, then $\delta \geq a_{\ell_2 +1} + 1$.
\end{enumerate}

\end{pro}

For $ 0 \leq \ell < z$, denote $\Delta(\ell) = \cup_{\nu=0}^\ell \mathcal{I}_{a_\nu}$ and  $\Delta(\ell)^{\perp_e} =H \setminus  \cup_{\nu=0}^\ell \mathcal{I}_{a_\nu}^R$. Then, by \cite[page 5]{GHR-RS}, the Euclidean dual of $\mathbb{F}_{p^s}^n \cap C_{\Delta(\ell)}$ coincides with the subfield-subcode $\mathbb{F}_{p^s}^n \cap C_{\Delta(\ell)^{\perp_e}}$. We decompose $\Delta(\ell) = \Delta_r \sqcup \Delta_L$, where $\Delta_r$ consists of the asymmetric cosets in $\Delta(\ell)$ whose reciprocal coset is not in $\Delta(\ell)$ and $\Delta_L$ is the union of the remaining asymmetric cosets and the symmetric ones. It is clear that $\Delta(\ell) = \Delta_r \cup \Delta_L$ and $ \Delta_r \cap \Delta_L = \emptyset$.

The following result follows by applying Theorem \ref{entang-css} (where one considers $C_1 = C_2 = \mathbb{F}_{p^s}^n \cap C_{\Delta(\ell)}$) and the first displayed formula in \cite[page 7]{GHR-RS}.

\begin{teo}
\label{BCH}
With the above notation, there exists an EAQECC whose parameters are
\[
\left[ \left[ n, n-2 \sum_{\nu=0}^{\ell} i_{a_\nu} + c, \geq a_{\ell +1} +1; c \right] \right]_{p^s},
\]
where $c = \# \Delta_L$, $\#$ meaning cardinality.
\end{teo}

The goal of this section is to compute the parameters of the Steane enlargement of some EAQECCs provided either by Theorem \ref{BCH} or by the same procedure as in that theorem but associated to another codes $C_\Delta$. Given a prime number $p$ and distinct positive integers $m$ and $s$ as above, we define the positive integer $B(p,m,s)$ as
\[
B(p,m,s) := \begin{cases}
\left( p^s \right)^{ \frac{m}{2s} } -1 & \text{ if } \frac{m}{s} \text{ is even},\\
\left( p^s \right)^{\lceil \frac{m}{2s} \rceil} -p^s +1  & \text{otherwise}.
\end{cases}
\]
We will use this integer frequently.

The following result follows from Theorem 3 and Lemma 8 in \cite{Aly}.

\begin{pro}
\label{bound}
Let $p, m$ and $s$ be integers as above and consider another integer $b$ such that $0 <b <B(p,m,s)$. Then,
\begin{description}
\item[a)] $\mathcal{I}_b \subseteq \mathcal{I}_b^{\perp_e} := H \setminus \mathcal{I}_b^R$.
 \item[b)] $\# \mathcal{I}_b = \frac{m}{s}$ and this equality also holds when $b = B(p,m,s)$.
\end{description}
\end{pro}






We are ready for stating our first new result in this section. It falls within the case described in Subsection \ref{sub:one}. We state it and afterwards we give two corollaries.

\begin{teo}
\label{BCHcase1}
Let $p$ be a prime number and $m$ and $s$ distinct positive integers as above. Set $n= p^m -1$ and $\mathcal{A}= \{a_0=0 < a_1 < \cdots < a_z\}$ the set of minimal representatives of the family of minimal cyclotomic cosets (with respect to $n$ and $s$). Pick indices $\ell_1 < \ell_2 < z$   such that $a_{\ell_2} < B(p,m,s)$. Then, there is a Steane enlargement of an  EAQECC as in Theorem \ref{BCH} whose parameters are
\[
\left[ \left [n, n - \frac{m}{s} (\ell_1 + \ell_2) -1, d';1 \right]\right]_{p^s},
\]
where $d' \geq \min \left\{ a_{\ell_2 +1} +1, \left\lceil \left(\frac{p^s+1}{p^s} \right) (a_{\ell_1 +1} +1) \right\rceil \right\}$.
\end{teo}

\begin{proof}
With the notation as in Theorem \ref{case1}, set $C := \mathbb{F}_{p^s} \cap C_{\Delta(\ell_2)}$ and $B_r$ (respectively, $B_t$) a generator matrix of the codes $\mathbb{F}_{p^s} \cap C_{\Delta(\ell_1)}$ (respectively, $\mathbb{F}_{p^s} \cap C_{\Delta'}$, where $\Delta' = \cup_{\nu=\ell_1 +1 }^{\ell_2} \mathcal{I}_{a_\nu}$). Then, taking into account that the evaluation of a monomial $X^a$ is orthogonal to that of $X^b$ except when $ a+b \equiv 0 \mod n$ (see \cite[Proposition 2.2]{QINP2}), by Proposition \ref{bound} one deduces that $\Delta_L =  \mathcal{I}_{0}$ in the decomposition $\Delta(\ell_2) = \Delta_r \sqcup \Delta_L$ given before Theorem \ref{BCH}. Thus the value $c$ of the EAQECC given by $C$ is $c=1$.
Finally, if one applies Theorem \ref{case1} and considers propositions \ref{grado2} and \ref{bound}, one obtains a Steane enlargement of the EAQECC given by $C$ with parameters
\[
\left[ \left [n, n - 2 \left( \frac{m \ell_1}{s} +1 \right) - \left( \frac{m (\ell_2 -\ell_1)}{s}\right) + 1, d';1 \right]\right]_{p^s},
\]
where $d' \geq \min \left\{ a_{\ell_2 +1} +1, \left\lceil \left(\frac{p^s+1}{p^s} \right) (a_{\ell_1 +1} +1) \right\rceil \right\}$. This concludes the proof.
\end{proof}

Taking $\ell_1=0$ and $\ell_2=1$ in the above theorem, one gets the following result.
\begin{cor}
\label{three}
Consider a prime number $p$ and a positive integer $n= p^m -1$ given by $m>0$. Set $s \neq m$ a positive integer such that $s$ divides $m$. Then, there exists a Steane enlargement of an EAQECC as in Theorem \ref{BCH} with parameters $\left[ \left[n,n- \frac{m}{s}  -1, 3;1 \right]\right]_{p^s}$.
\end{cor}

The following result is a bit weaker than Theorem \ref{BCHcase1} but depends only on an element in $\mathcal{A}$.
\begin{cor}
\label{one}
Keep the notation as in Theorem \ref{BCHcase1} and pick $a_{\ell} \in \mathcal{A}$ such that $a_{\ell} < B(p,m,s)$. Then, there is a Steane enlargement of an EAQECC as in Theorem \ref{BCH} whose parameters are
\[
\left[ \left [n, n - \frac{m}{s} ( 2 \ell -1 ) -1, a_{\ell +1} ;1 \right]\right]_{p^s}.
\]
\end{cor}
\begin{proof}
Apply Theorem \ref{BCHcase1} for $\ell_1 = \ell -1$ and $\ell_2 = \ell$. The proof follows from the fact that $a_{\ell +1} = a_{\ell} +2$ if $a_\ell$ +1 is a multiple of $p^s$ and $a_{\ell +1} = a_{\ell} +1$ otherwise. Indeed, it is clear that $a_i=i$ whenever $i < p^s$ and that, when $a_\ell +1$ is a multiple of $p^s$, then $a_{\ell +1} > a_\ell +1$. Thus, when $m/s$ is even, it suffices to prove that the elements of the form $b + \lambda p^s < p^{m/2} -1$, $b, \lambda$ positive integers and $0 \neq b < p^s$ are minimal representatives in $\mathcal{A}$. This is true because $b$ is the first element in the $p^s$-adic expansion of $b + \lambda p^s$ and, expressing the $p^s$-adic expansion $a_0 + a_1 p^s + \cdots + a_{\lfloor \frac{n}{s}\rfloor} p^{\lfloor \frac{n}{s}\rfloor s}$ of an integer as $(a_0, a_1, \ldots, a_{\lfloor \frac{n}{s}\rfloor})$,  the $p^s$-adic expansions of the elements in the coset $ \mathcal{I}_{b + \lambda p^s}$ are obtained by successively shifting the $p^s$-adic expansion of $b + \lambda p^s$. During a while, these shifts are $p^s$-adic expansions with a zero in the first position and when one obtains a nonzero in the first position, the corresponding value is larger than $p^{m/2}-1$. An analogous reasoning proves the result in the case when $m/2$ is odd.
\end{proof}

The following result is also supported on Theorem \ref{case1}. We keep the same conditions as in Theorem \ref{BCHcase1}. That is, we consider a prime number $p$, and $m$ and $s$ different positive integers such that $s$ divides $m$. Let $\mathcal{A}= \{a_0=0 < a_1 < \cdots < a_z\}$ be the set of minimal cyclotomic cosets (with respect to $n = p^m -1$ and $s$).

\begin{teo}
\label{Reciprocal}
Let $\ell_1 < \ell_2 < z$ two indices such that $a_{\ell_2} < B(p,m,s)$. Then, there is a Steane enlargement of  an EAQECC determined by a code $C_\Delta$ with parameters:
\[
\left[ \left [n, n - \frac{m}{s} (\ell_1 + \ell_2) -1, d';c \right]\right]_{p^s},
\]
where $d' \geq \min \left\{ a_{\ell_1 +1} + a_{\ell_2 +1}, \left\lceil 2 \left(\frac{p^s+1}{p^s} \right) a_{\ell_1 +1} \right\rceil \right\}$ and $c=1 + 2 \frac{m}{s} \ell_1$.
\end{teo}
\begin{proof}
For a nonnegative integer $\ell$, define $\Delta(\ell, R) : = \cup_{\nu=0}^{\ell} \mathcal{I}_{a_\nu} \bigcup \cup_{\nu=1}^{\ell} \mathcal{I}_{a_\nu}^R$. Now, with the notation as in Theorem \ref{case1}, set $C := \mathbb{F}_{p^s} \cap C_{\Delta(\ell_1,R) \cup \Delta'}$, where $\Delta' = \cup_{\nu= \ell_1 +1}^{\ell_2} \mathcal{I}_{a_\nu}$. Fix $B_r$ (respectively, $B_t$) a generator matrix of the code $\mathbb{F}_{p^s} \cap C_{\Delta(\ell_1, R)}$ (respectively, $\mathbb{F}_{p^s} \cap C_{\Delta'}$).

To compute $c$, we reason as in the proof of Theorem \ref{BCHcase1} and $c=1 + 2 \frac{m}{s} \ell_1$, because the set $\Delta_L$ in the decomposition $\Delta(\ell_1,R) \cup \Delta' = \Delta_r \sqcup \Delta_L$ given before Theorem \ref{BCH} is $\Delta(\ell_1,R)$. With respect to the distance, we notice that the values in $\Delta(\ell, R)$ contain all the consecutive integers from 0 to $a_{\ell +1} -1$ and their opposites modulo $n$. Then, using the $*$ product defined by $(x_1, \ldots, x_n)*(y_1, \ldots, y_n) =(x_1 y_1, \ldots, x_n y_n)$, we deduce that there is code which is isometric to $C_{\Delta(\ell, R)}$ containing the evaluation of consecutive powers of $X$. This proves that the minimum Hamming distance of $\langle B_r \rangle^{\perp_e}$ is larger than or equal to $2 a_{\ell_1 +1}$. A close reasoning shows that $d_H(C^{\perp_e}) \geq a_{\ell_1 +1} + a_{\ell_2 +1}$, which ends the proof.
\end{proof}

Our last results fit with the case described in Subsection \ref{sub:two}. We start with the following one.

\begin{teo}
\label{BCHcase2}
Let $p$ be a prime number and $m$ and $s$ positive integers such that $s$ divides $m$, $s\neq m$. Assume that $m$ and $m/s$ are even. Set $n= p^m -1$ and $\mathcal{A}= \{a_0=0 < a_1 < \cdots < a_z\}$ the set of minimal representatives of the family of minimal cyclotomic cosets (with respect to $n$ and $s$). Let $0 < \ell_2 <z$ that index such that $a_{\ell_2} = p^{\frac{m}{2}}-1$. Consider an index $ 0 \leq \ell_1 < \ell_2$. Then, there is a Steane enlargement of an  EAQECC as in Theorem \ref{BCH} whose parameters are
\[
\left[ \left [n, n - \frac{m}{s} (\ell_1 + \ell_2) -1 + \frac{m}{2s}, d'; \frac{m}{2s}+1 \right]\right]_{p^s},
\]
where $d' \geq \min \left\{ a_{\ell_2 +1} +1, \left\lceil \left(\frac{p^s+1}{p^s} \right) (a_{\ell_1 +1} +1) \right\rceil \right\}$.
\end{teo}
\begin{proof}
We desire to use Theorem \ref{case2}. Consider $C := \mathbb{F}_{p^s} \cap C_{\Delta(\ell_2)}$ and a suitable generator matrix $B_r$ (respectively, $B_t$) of the subfield-codes $\mathbb{F}_{p^s} \cap C_{\Delta(\ell_1)}$ (respectively, $\mathbb{F}_{p^s} \cap C_{\Delta'}$, where $\Delta' = \cup_{\nu=\ell_1 +1 }^{\ell_2} \mathcal{I}_{a_\nu}$). With the notation used after Proposition \ref{grado2} and by \cite[Remark 3.4]{QINP2}, it holds that $\Delta_L =  \mathcal{I}_{0} \cup \mathcal{I}_{a_{\ell_2}}$ and $\Delta_r = \cup_{\nu=1}^{\ell_{2}-1} \mathcal{I}_{a_\nu}$ in  the decomposition $\Delta(\ell_2) = \Delta_r \sqcup \Delta_L$. Theorem \ref{BCH} and Proposition \ref{bound} show that the value $c$ of the EAQECC given by $C$ is $c = 1 + \# \mathcal{I}_{a_{\ell_2}} =1 + (m/s)$.

Finally, the proof follows by applying Theorem \ref{case2} and Corollary \ref{matrix} after noticing that $m/s$ is the rank of a suitable matrix $A_0 - A_0^T$ as introduced in Subsection \ref{sub:two} and $\mathrm{rank}(B_r B_r^T) =1$. Thus we have proved that there is a Steane enlargement of the EAQECC given by $C$ with parameters as in the statement.

\end{proof}

As in the proof of Theorem \ref{Reciprocal}, we can add reciprocal cosets and the obtained result is the following one.

\begin{teo}
\label{Reciprocal-2} The following statements hold.
\begin{description}
  \item[i)] Keep the same notation and  requirements as in Theorem \ref{Reciprocal}, then there exists a Steane enlargement of  an EAQECC determined by a code $C_\Delta$ with parameters:
\[ \left[ \left [n, n - \frac{m}{s} (\ell_1 + \ell_2) -1, d';c' \right]\right]_{p^s},\]
where $d' \geq \min \left\{ 2 a_{\ell_2 +1}, \left\lceil 2 \left(\frac{p^s+1}{p^s} \right) a_{\ell_1 +1} \right\rceil \right\}$ and $c'=1 +  \frac{m}{s} (\ell_1+ \ell_2)$.
  \item[ii)] If we are under the notation and conditions of Theorem \ref{BCHcase2}, then there is a Steane enlargement of  an EAQECC with parameters
\[
\left[ \left [n, n - \frac{m}{s} \left(\ell_1 + \ell_2 - \frac{1}{2}\right) -1, d';c' \right]\right]_{p^s},
\]
where $$d' \geq \min \left\{ a_{\ell_2 +1} + a_{\ell_2} -1, \left\lceil 2 \left(\frac{p^s+1}{p^s} \right) a_{\ell_1 +1} \right\rceil \right\}$$ and $c'=1 + \frac{m}{s} (\ell_1 + \ell_2 - \frac{1}{2})$.
\end{description}
\end{teo}

\begin{proof}
To prove the first statement, it suffices to consider $C := \mathbb{F}_{p^s} \cap C_{\Delta(\ell_2,R)}$ and suitable matrices $B_r$ and $B_t$, where $B_r$ generates the subfield-code $\mathbb{F}_{p^s} \cap C_{\Delta(\ell_1,R)}$ and, then, apply the same procedure as in the proof of Theorem \ref{BCHcase2}. A proof for the second statement is analogous after taking into account that the coset $\mathcal{I}_{a_{\ell_2}}$ is symmetric. Recall that in this last case  $a_{\ell_2} = p^{\frac{m}{2}}-1$.
\end{proof}

\begin{rem}
\label{La22}
{\rm
Under the conditions described in the second paragraph of Remark \ref{La13}, we can also obtain Steane enlargements of EAQECCs whose parameters coincide with those stated in Theorems \ref{BCHcase2} and \ref{Reciprocal-2} but the entanglement $c'$ and the dimension, which are one unit less. To do it, it suffices to consider matrices $A_0$ defined by polynomials $\eta_{t_\ell}$ as in the mentioned Remark \ref{La13}.
}
\end{rem}



\subsection{Examples}
\label{examples}
We conclude this section and the paper by providing parameters of EAQECCs obtained from some results given in Subsection \ref{results}. Note that Steane enlarged codes enjoy the interesting computational advantages described after Theorem \ref{CSS}. We present some EAQECCs with parameters that we have not found in the literature and cannot be obtained from existing ones by means of propagation rules.



We give three tables of $q$-ary EAQECCs for $q=2, 4, 9$. Our tables show the parameters of the codes and the involved results to deduce them. The tables also present the values $m$, $s$, $a_{\ell_1}$ and $a_{\ell_1}$ used in our computations.

Table \ref{ta:uno} shows some examples in the binary case. To enhance understanding we explain three cases in detail. The same procedure allows us to obtain the parameters of all our tables. The first binary code in the table have parameters $[[15,4,6;3]]_2$ and them follow from the formula in Theorem \ref{BCHcase2}, where $m=4$, $s=1$, $\ell_1=1$, $a_{\ell_1}= 1$, $\ell_2=2$, $a_{\ell_2}= 3$ and $a_{3}= 5$. If one applies Theorem \ref{BCHcase1} for $m=5$, $s=1$, $\ell_1=1$, $a_{\ell_1}= 1$, $\ell_2=2$, $a_{\ell_2}= 3$ and $a_{3}= 5$, then one gets a code with parameters $[[31,15,6;1]]_2$ as in the second line of the table. Finally, the binary code $[[63,32,14;31]]_2$ in the eighth row of the table can be constructed from Theorem \ref{Reciprocal-2} by noticing that $m=6$, $s=1$, $\ell_1=2$, $a_{\ell_1}= 3$, $\ell_2=3$, $a_{\ell_2}= 5$ and $a_{4} =7$. To the best of our knowledge, EAQECCs in Table \ref{ta:uno} are new with the exception of those marked with a *. We think they are good since in some cases they compare well with others in the literature. Indeed, our code $[[15,4,6;3]]_2$ (respectively, $[[63,44,8;13]]_2$)  is better than $[[15,4,6;5]]_2$ (respectively, $[[63,42,8;14]]_2$) appearing in \cite{codetables}. This last reference also contains the codes marked with a *, being the best known codes with those parameters. We add them to show that we are able to get good codes.

\begin{table}
\begin{center}
\begin{tabular}{|c|c|c|c||c|c|c|c|}
  \hline
  $n$ & $k$ & $d\ge$  & $c$ & Result & $(m,s)$ & $a_{\ell_1}$ & $a_{\ell_2}$ \\
  \hline
  15 & 4 & 6 & 3 & Theorem \ref{BCHcase2} & (4,1) & 1 & 3 \\
  31 & 15 & 6 & $1^*$ & Theorem \ref{BCHcase1} & (5,1) & 1 & 3 \\
  31 & 15 & 8 & 11 & Theorem \ref{Reciprocal} & (5,1) & 1 & 3 \\
  63 & 44 & 8 & 13 & Theorem \ref{Reciprocal} & (6,1) & 1 & 3 \\
  63 & 32 & 8 & $1^*$  & Theorem \ref{BCHcase1} & (6,1) & 3 & 5 \\
  63 & 44 & 9 & 19 & Theorem \ref{Reciprocal-2} & (6,1) & 1 & 3 \\
  63 & 38 & 9 & 13 & Theorem \ref{Reciprocal} & (6,1) & 1 & 5 \\
  63 & 32 & 14 & 31  & Theorem \ref{Reciprocal-2} & (6,1) & 3 & 5 \\
  127 & 70 & 12 & 1 & Theorem \ref{BCHcase1} & (7,1) & 5 & 9 \\
  127 & 84 & 14 & 29 & Theorem \ref{Reciprocal} & (7,1) & 3 & 7 \\
  127 & 70 & 18 & 43 & Theorem \ref{Reciprocal} & (7,1) & 5 & 7 \\
  255 & 214 & 12 & 33 & Theorem \ref{Reciprocal} & (8,1) & 3 & 5 \\
  255 & 214 & 14 & 41 & Theorem \ref{Reciprocal-2} & (8,1) & 3 & 5 \\
  255 & 190 & 18 & 49 & Theorem \ref{Reciprocal} & (8,1) & 5 & 9 \\
  \hline
\end{tabular}
\caption{Parameters of binary  EAQECCs }\label{ta:uno}
\end{center}
\end{table}

Table \ref{ta:dos} (respectively, \ref{ta:tres}) provides some examples of new $4$-ary (respectively, $9$-ary) EAQECCs obtained with our results. Furthermore, according to Remark \ref{La22}, the values $k$ and $c$ can be decreased one unit whenever we apply Theorems \ref{BCHcase2} and \ref{Reciprocal-2}. Indeed, one can use the polynomials $\eta_{t_\ell}$ given in Remark \ref{La13} after noticing that $t_\ell = 6$ in the corresponding cases of Table \ref{ta:tres}.

\begin{table}
\begin{center}
\begin{tabular}{|c|c|c|c||c|c|c|c|}
  \hline
  $n$ & $k$ & $d\ge$  & $c$ & Result & $(m,s)$ & $a_{\ell_1}$ & $a_{\ell_2}$ \\
  \hline
  63 & 59 & 3 & 1 & Theorem \ref{BCHcase1} & (6,2) & 0 & 1 \\
  63 & 26 & 17 & 31 & Theorem \ref{Reciprocal} & (6,2) & 6 & 9 \\
  63 & 23 & 19 & 37 & Theorem \ref{Reciprocal} & (6,2) & 7 & 9 \\
  63 & 23 & 20 & 40 & Theorem \ref{Reciprocal-2} & (6,2) & 7 & 9 \\
  63 & 14 & 23 & 43 & Theorem \ref{Reciprocal} & (6,2) & 9 &11 \\
  63 & 11 & 26 & 52 & Theorem \ref{Reciprocal-2} & (6,2) & 10 & 11 \\
  1023 & 1017 & 3 & 1 & Theorem \ref{BCHcase1} & (10,2) & 0 & 1 \\
  1023 & 1007 & 4 & 1 & Theorem \ref{BCHcase1} & (10,2) & 1 & 2 \\
  1023 & 892 & 36 & 121 & Theorem \ref{Reciprocal} & (10,2) & 15 & 18 \\
  1023 & 887 & 37 & 131 & Theorem \ref{Reciprocal} & (10,2) & 17 & 18 \\
  1023 & 887 & 38 & 136 & Theorem \ref{Reciprocal-2} & (10,2) & 17 & 18 \\
  1023 & 877 & 40 & 141 & Theorem \ref{Reciprocal} & (10,2) & 18 & 19 \\
  1023 & 877 & 42 & 146 & Theorem \ref{Reciprocal-2} & (10,2) & 18 & 19 \\
  1023 & 847 & 50 & 176 & Theorem \ref{Reciprocal-2} & (10,2) & 22 & 23 \\
  1023 & 602 & 113 & 411 & Theorem \ref{Reciprocal} & (10,2) & 54 & 57 \\
  1023 & 597 & 115 & 421 & Theorem \ref{Reciprocal} & (10,2) & 55 & 57 \\
\hline
\end{tabular}
\caption{Parameters of $4$-ary  EAQECCs }\label{ta:dos}
\end{center}
\end{table}

\begin{table}
\begin{center}
\begin{tabular}{|c|c|c|c||c|c|c|c|}
  \hline
  $n$ & $k$ & $d\ge$  & $c$ & Result & $(m,s)$ & $a_{\ell_1}$ & $a_{\ell_2}$ \\
  \hline
  728 & 724 & 3 & 1 & Theorem \ref{BCHcase1} & (6,2) & 0 & 1 \\
 728 & 526 & 78 & 202 & Theorem \ref{Reciprocal-2} & (6,2) & 37 & 38 \\
 728 & 517 & 80 & 205 & Theorem \ref{Reciprocal} & (6,2) & 38 & 40 \\
 728 & 514 & 81 & 211 & Theorem \ref{Reciprocal} & (6,2) & 39 & 40 \\
 728 & 511 & 82 & 211 & Theorem \ref{Reciprocal} & (6,2) & 39 & 41 \\
728 & 508 & 83 & 217 & Theorem \ref{Reciprocal} & (6,2) & 40 & 41 \\
 728 & 496 & 88 & 232 & Theorem \ref{Reciprocal-2} & (6,2) & 42 & 43 \\
 728 & 490 & 90 & 235 & Theorem \ref{Reciprocal} & (6,2) & 43 & 44 \\
 728 & 490 & 92 & 238 & Theorem \ref{Reciprocal-2} & (6,2) & 43 & 44 \\
\hline
\end{tabular}
\caption{Parameters of $9$-ary  EAQECCs }\label{ta:tres}
\end{center}
\end{table}

\section*{Conflict of interest}
The authors declare they have no conflict of interest.

\bibliography{biblioEA}

\begin{thebibliography}{10}

\bibitem{Aly}
S.A. Aly, A.~Klappenecker, and P.~K. Sarvepalli.
\newblock On quantum and classical {BCH} codes.
\newblock {\em IEEE Trans. Inform. Theory}, 53(3):1183--1188, 2007.

\bibitem{Aru}
F.~Arute~et al.
\newblock Quantum supremacy using a programmable superconducting processor.
\newblock {\em Nature}, 574:505--510, 2019.

\bibitem{AK}
A.~Ashikhmin and E.~Knill.
\newblock Nonbinary quantum stabilizer codes.
\newblock {\em IEEE Trans. Inform. Theory}, 47(7):3065--3072, 2001.

\bibitem{7kkk}
A.E. Ashikhmin, A.M. Barg, E.~Knill, and S.N. Litsyn.
\newblock Quantum error detection {I}. {S}tatement of the problem.
\newblock {\em IEEE Trans. Inform. Theory}, 46(3):778--788, 2000.

\bibitem{8kkk}
A.E. Ashikhmin, A.M. Barg, E.~Knill, and S.N. Litsyn.
\newblock Quantum error detection {II}. {B}ounds.
\newblock {\em IEEE Trans. Inform. Theory}, 46(3):789--800, 2000.

\bibitem{Bier}
J.~Bierbrauer.
\newblock The theory of cyclic codes and a generalization to additive codes.
\newblock {\em Des. Codes Cryptogr.}, 25(2):189--206, 2002.

\bibitem{BE}
J.~Bierbrauer and Y.~Edel.
\newblock Quantum twisted codes.
\newblock {\em J. Comb. Designs}, 8:174--188, 2000.

\bibitem{Brun}
T.~Brun, I.~Devetak, and M.~Hsieh.
\newblock Correcting quantum errors with entanglement.
\newblock {\em Science}, 314(5798):436--439, 2006.

\bibitem{Calderbank}
A.~R. Calderbank, E.~M. Rains, P.~W. Shor, and N.~J.~A. Sloane.
\newblock Quantum error correction via codes over {${\rm GF}(4)$}.
\newblock {\em IEEE Trans. Inform. Theory}, 44(4):1369--1387, 1998.

\bibitem{20kkk}
A.~R. Calderbank and Peter~W. Shor.
\newblock Good quantum error-correcting codes exist.
\newblock {\em Phys. Rev. A}, 54:1098--1105, Aug 1996.

\bibitem{cao-cui2}
M.~Cao and J~Cui.
\newblock New stabilizer codes from the construction of dual-containing
  matrix-product codes.
\newblock {\em Finite Fields Appl.}, 63:101643, 2020.

\bibitem{Cas}
I.~Cascudo.
\newblock On squares of cyclic codes.
\newblock {\em IEEE Trans. Inform. Theory}, 65(2):1034--1047, 2019.

\bibitem{Chen2}
X.~Chen, S.~Zhu, and X.~Kai.
\newblock Entanglement-assisted quantum negacyclic {BCH} codes.
\newblock {\em Internat. J. Theoret. Phys.}, 58(5):1509--1523, 2019.

\bibitem{Delgo}
P.~Delsarte and J.~M. Goethals.
\newblock Alternating bilinear forms over {GF}(q).
\newblock {\em J. Comb. Th.}, 19:26--50, 1975.

\bibitem{Mesnager}
Z.~Du, C.~Li, and S.~Mesnager.
\newblock Constructions of self-orthogonal codes from hulls of {BCH} codes and
  their parameters.
\newblock {\em IEEE Trans. Inform. Theory}, pages 6774--6785, 2020.

\bibitem{QINP2}
C.~Galindo, O.~Geil, F.~Hernando, and D.~Ruano.
\newblock On the distance of stabilizer quantum codes from {$J$}-affine variety
  codes.
\newblock {\em Quantum Inf. Process.}, 16(4):Art. 111, 32, 2017.

\bibitem{gahe}
C.~Galindo and F.~Hernando.
\newblock On the generalization of the construction of quantum codes from
  {H}ermithian self-orthogonal codes.
\newblock {\em Des. Codes Cryptogr.}, 90:1103--1112, 2022.

\bibitem{Martin-1}
C.~Galindo, F.~Hernando, H.~Mart\'{\i}n-Cruz, and D.~Ruano.
\newblock Stabilizer quantum codes defined by trace-depending polynomials.
\newblock {\em Finite Fields Appl.}, 87:102138, 2023.

\bibitem{QINP3}
C.~Galindo, F.~Hernando, R.~Matsumoto, and D.~Ruano.
\newblock Entanglement-assisted quantum error-correcting codes over arbitrary
  finite fields.
\newblock {\em Quantum Inf. Process.}, 18(4):Art. 116, 18, 2019.

\bibitem{QINP}
C.~Galindo, F.~Hernando, and D.~Ruano.
\newblock Stabilizer quantum codes from {$J$}-affine variety codes and a new
  {S}teane-like enlargement.
\newblock {\em Quantum Inf. Process.}, 14(9):3211--3231, 2015.

\bibitem{GHR-RS}
C.~Galindo, F.~Hernando, and D.~Ruano.
\newblock Entanglement-assisted quantum error-correcting from {RS} codes and
  {BCH} codes with extension degree 2.
\newblock {\em Quantum Inf. Process.}, 20:158, 2021.

\bibitem{Gottesman}
D.~Gottesman.
\newblock Class of quantum error-correcting codes saturating the quantum
  {H}amming bound.
\newblock {\em Phys. Rev. A}, 54(3):1862--1868, 1996.

\bibitem{FTGot}
D.~Gottesman.
\newblock Fault-tolerant quantum computation with higher-dimensional systems.
\newblock In Colin~P. Williams, editor, {\em Quantum Computing and Quantum
  Communications}, pages 302--313, Berlin, Heidelberg, 1999. Springer Berlin
  Heidelberg.

\bibitem{Grassl22}
M.~Grassl.
\newblock New quantum codes from {CSS} codes.
\newblock {\em Quantum Inf. Process.}, 22:86, 2023.

\bibitem{codetables}
M.~Grassl.
\newblock Bounds on the minimum distance of linear codes.
\newblock {\em www.codetables.de}, accessed on 22 November 2022.

\bibitem{Guo}
G.~Guo and R.~Li.
\newblock New entanglement-assisted quantum {MDS} codes derived from
  generalized {R}eed-{S}olomon codes.
\newblock {\em Internat. J. Theoret. Phys.}, 59(4):1241--1254, 2020.

\bibitem{Hamada}
M.~Hamada.
\newblock Concatenated quantum codes constructible in polynomial time:
  Efficient decoding and error correction.
\newblock {\em IEEE Trans. Inform. Theory}, 54(12):5689--5704, 2008.

\bibitem{Hsie}
M.~H. Hsieh, I.~Devetak, and T.~Brun.
\newblock General entanglement-assisted quantum error-correcting codes.
\newblock {\em Phys. Rev. A}, 76:062313, Dec 2007.

\bibitem{Ioffe}
L.~Ioffe and M.~M\'ezard.
\newblock Asymmetric quantum error-correcting codes.
\newblock {\em Phys. Rev. A}, 75:032345, Mar 2007.

\bibitem{Ketkar}
A.~Ketkar, A.~Klappenecker, S.~Kumar, and P.~K. Sarvepalli.
\newblock Nonbinary stabilizer codes over finite fields.
\newblock {\em IEEE Trans. Inform. Theory}, 52(11):4892--4914, 2006.

\bibitem{FTKnill}
E.~Knill, R.~Laflamme, and W.H. Zurek.
\newblock Resilient quantum computation: {E}rror models and thresholds.
\newblock {\em Proc. Royal Soc. London A}, 454:365--384, 1998.

\bibitem{Lag2}
G.~G. La~Guardia.
\newblock On the construction of nonbinary quantum bch codes.
\newblock {\em IEEE Trans. Inform. Theory}, 60(3):1528--1535, 2014.

\bibitem{Chao}
S.~Ling, J.~Luo, and C.~Xing.
\newblock Generalization of {S}teane's enlargement construction of quantum
  codes and applications.
\newblock {\em IEEE Trans. Inform. Theory}, 56(8):4080--4084, 2010.

\bibitem{Guo2}
G.~Luo and X.~Cao.
\newblock Two new families of entanglement-assisted quantum {MDS} codes from
  generalized {R}eed-{S}olomon codes.
\newblock {\em Quantum Inf. Process.}, 18(3):Paper No. 89, 12, 2019.

\bibitem{LuoEz}
G.~Luo, M.F. Ezerman, and S.~Ling.
\newblock Entanglement-assisted and subsystem quantum codes. {N}ew propagation
  rules and constructions.
\newblock {\em arXiv:2206.09782}, 2022.

\bibitem{luol}
L.~Luo and Z.~Ma.
\newblock Fault-tolerant quantum computation with non-binary systems.
\newblock {\em Quantum Inf. Process.}, 18:188, 2019.

\bibitem{Qian}
J.~Qian and L.~Zhang.
\newblock On {MDS} linear complementary dual codes and entanglement-assisted
  quantum codes.
\newblock {\em Des. Codes Cryptogr.}, 86(7):1565--1572, 2018.

\bibitem{Quian2}
J.~Qian and L.~Zhang.
\newblock Constructions of new entanglement-assisted quantum {MDS} and almost
  {MDS} codes.
\newblock {\em Quantum Inf. Process.}, 18(3):Paper No. 71, 12, 2019.

\bibitem{Diego}
D.~Ruano.
\newblock The metric structure of linear codes.
\newblock In {\em Singularities, algebraic geometry, commutative algebra, and
  related topics}, pages 537--561. Springer, Cham, 2018.

\bibitem{Sari}
M.~Sari and E.~Kolotoglu.
\newblock An application of constacyclic codes to entanglement-assisted quantum
  {MDS} codes.
\newblock {\em Comput. Appl. Math.}, 38(2):Paper No. 75, 13, 2019.

\bibitem{Ass}
P.~K. Sarpevalli, A.~Klappenecker, and M.~Rotteler.
\newblock Assymmetric quantum codes: constructions, bounds and perfomance.
\newblock {\em Proc. Roy. Soc. A}, 465:1645--1672, 2000.

\bibitem{Shor-Preskill}
P.~W. Shor and J.~Preskill.
\newblock Simple proof of security of the {BB84} quantum key distribution
  protocol.
\newblock {\em Phys. Rev. Lett.}, 85:441--444, Jul 2000.

\bibitem{FTShor}
P.W. Shor.
\newblock Fault-tolerant quantum computation.
\newblock {\em Proc. 37th ann. symp. found. comp. sc., IEEE Comp. Soc. Press},
  pages 56--65, 1996.

\bibitem{95kkk}
A.~Steane.
\newblock Simple quantum error correcting codes.
\newblock {\em Phys. Rev. Lett.}, 77:2551--2577, 1996.

\bibitem{Steane-enl}
A.M. Steane.
\newblock Enlargement of {C}alderbank-{S}hor-{S}teane quantum codes.
\newblock {\em IEEE Trans. Inform. Theory}, 45(7):2492--2495, 1999.

\bibitem{Wilde}
M.~M. Wilde and T.~A. Brun.
\newblock Optimal entanglement formulas for entanglement-assisted quantum
  coding.
\newblock {\em Phys. Rev. A}, 77:064302, Jun 2008.

\end{thebibliography}
\bibliographystyle{plain}

\end{document}